\documentclass[%
 reprint,
 amsmath,amssymb,
 aps,
]{revtex4-2}

\usepackage{graphicx}
\usepackage{dcolumn}
\usepackage{bm}

\usepackage[final]{hyperref}
\usepackage{float}
\hypersetup{
	colorlinks=true,       
	linkcolor=blue,        
	citecolor=blue,        
	filecolor=magenta,     
	urlcolor=blue         
}
\usepackage{braket}
\usepackage{cleveref}
\usepackage{verbatim}
\usepackage{ulem}

\begin{document}

\preprint{APS/123-QED}

\title{Classical Larmor formula through the Unruh effect for uniformly accelerated electrons}

\author{Georgios Vacalis$^{1}$}
\thanks{Corresponding author: georgios.vacalis@chch.ox.ac.uk}
\author{Atsushi Higuchi$^{2}$}
\author{Robert Bingham$^{3,4}$}
\author{Gianluca Gregori$^{1}$}

\affiliation{ \small\it $^1$Department of Physics, University of Oxford, Parks Road, Oxford OX1 3PU, UK}
\affiliation{ \small\it$^2$ Department of Mathematics, University of York, Heslington, York YO10 5DD, UK}
\affiliation{\small\it $^3$Rutherford Appleton Laboratory, Chilton, Didcot, Oxon OX11 0QX, UK,}%

\affiliation{\small\it $^4$Department of Physics, University of Strathclyde, Glasgow G4 0NG, UK}%
\date{\today}

\begin{abstract}
We investigate the connection between the classical Larmor formula and the quantum Unruh effect by computing the emitted power by a 
uniformly accelerated charged particle and its angular distribution in the co-accelerated frame. We consider a classical particle accelerated
with non-zero charge only for a finite period and then take the infinite-time limit after removing the effects due to the initial charging
and final discharging processes. 
We show that the result found for the interaction rates agrees with previous studies in 
which the period of acceleration with non-zero charge was taken to be infinite from the beginning.  
We also show that the power and angular distribution
of emission, which is attributed either to the emission or absorption of a Rindler photon in the co-accelerated frame, is given by 
the Larmor formula, confirming that, at tree level, it is necessary to take into account the Unruh effect in order to reproduce the 
classical Larmor radiation formula in the co-accelerated frame.
\end{abstract}

\maketitle


\section{Introduction} 

It is well established that spontaneous particle production can occur in curved spacetime \cite{Bire82,Fulling:1989nb,Parker:2009uva}. 
This effect has played a significantly r\^ole in our understanding of the early universe \cite{PhysRevLett.21.562,PhysRev.183.1057,Parker:2012at}. For example, gravitational particle production can provide a generation mechanism for dark matter (DM) especially during the inflationary period because of the high Hubble rate and curvature of spacetime (see Refs.~\cite{Klaric:2022qly,Ema:2019yrd,Cembranos:2019qlm,kolb2023cosmological} and references therein for recent studies involving DM candidates of different spin). Depending on the mass of the dark particle, this production channel could account for all observed DM in the universe.
Another well-recognized result of gravitationally induced particle production is the thermal radiation occurring near the event horizon of a black hole, known as Hawking radiation \cite{1974Natur.248...30H,Hawking:1975vcx}. Shortly after this discovery, it was shown  that a uniformly accelerated detector in flat spacetime also sees a thermal bath with temperature proportional to its own acceleration. This is known as the Fulling-Davies-Unruh (FDU) effect, or simply referred to as Unruh radiation \cite{PhysRevD.7.2850,Davies:1974th,PhysRevD.14.870}. Hawking and Unruh radiations are connected through the equivalence principle. 

So far, there has been no direct measurement of gravitational particle production, including, in particular, Hawking radiation. On the other hand, exploiting the equivalence principle, it would appear that measuring Unruh radiation from accelerating bodies is, by
itself, a test of gravitational particle production. An experiment has been proposed to verify the existence of the FDU bath which is encoded in Larmor radiation \cite{Cozzella:2017ckb}. Getting the required accelerations to produce an observable effect is challenging \cite{Crispino2008, gregori2023measuring}.  In this endeavor, the electron is the simplest `detector' to test the Unruh predictions. In fact, large accelerations can already be realised in the laboratory using high-intensity lasers \cite{WeberS,Gales_2018,Shen_2018} corresponding to a thermal bath of temperature $\gtrsim 1$ eV \cite{Chen:1998kp}. Additionally, radiation reaction becomes important for 1 micron lasers with intensities around $10^{21}$ W$\cdot$cm$^{-2}$ \cite{Brodin:2007vf}. While, in principle, this can be measured in the laboratory, detection of Unruh radiation has been a controversial subject in the literature \cite{Narozhny.2002,Ford:2005sh, Cruz2016}, especially concerning how to distinguish it from other classical and quantum radiation processes involving acceleration of charged particles.

In this paper, we aim at providing some clarification of this issue and show, using a full quantum field theory calculation, that the Unruh effect involving an accelerated electron reduces, at tree level, to nothing other than the classical Larmor radiation as seen in the laboratory frame. While this result would seem to diminish the importance of the FDU effect, 
it has, on the contrary, a fundamental implication.
Hawking's derivation of gravitational particle production makes use of untested approximations in which the appearance of trans-Planckian frequencies is unavoidable. 
The same problem is also present in Unruh's derivation. The fact that the FDU effect on an accelerated electron reduces to the experimentally-verified Larmor radiation gives strong support to Hawking's derivation of gravitational particle production. 

The presence of the FDU thermal bath is necessary when comparing photon emission rates in the inertial and and co-accelerated frame. In this context it is important to note that absorption of a photon in the FDU thermal bath in the co-accelerated frame corresponds to emission of a photon in the inertial frame \cite{PhysRevD.29.1047,PhysRevD.38.1118}.  It was shown in Refs.~\cite{PhysRevD.45.R3308,Higuchi:1992td} that the rate of photon emission by an accelerated charge in the inertial (or Minkowski) frame is the same as the sum of the rates of emission and absorption of photons in the co-accelerated or Rindler frame in the presence of a FDU thermal bath. In Ref.~\cite{Paithankar:2020akh}, the same equivalence was demonstrated for a more general case, where the uniformly accelerated charge has an arbitrary transverse motion. This connection suggests that the classical Larmor radiation can be seen as a consequence of the FDU thermal bath though the link between the two seems counter-intuitive since the former is a classical effect while the latter is a purely quantum one. In Refs.~\cite{Ren1994, PhysRevD.100.045020}, it was shown that the classical radiation is built from zero-energy Rindler modes, and the Larmor formula is recovered in the Rindler frame by coupling a scalar field to the accelerated particle. In Ref.~\cite{Lynch:2019hmk} the Larmor formula was recovered in Minkowski frame for photons instead of scalars. 

To fully clarify how the Unruh effect on an accelerated charged particle reduces to Larmor radiation, however, what has been missing in the above work is a calculation of the total photon power emitted by the accelerated electron using the Unruh effect in the Rindler frame. This task is carried out in this paper. The calculation will be done at tree level in the Rindler frame. Next-to-leading-order Feynman diagrams such as photon scattering by the electron (i.e., Compton scattering) are purely quantum, meaning that they have no classical equivalent. These higher-order processes are also linked to the Unruh effect
\cite{Sch_tzhold_2009}, but provide a much smaller contribution to the total
power \cite{PhysRevLett.97.121302,PhysRevLett.100.091301}. In this paper, we will
neglect any sub-dominant terms and leave them to a future study. In what follows, we use the metric signature $(+,-,-,-)$ and natural units $\hbar = c =  k_B =1$, unless stated otherwise. 

\section{Minkowski and Unruh modes}
\label{rinun} 

The goal of this section is to find the relation between the Minkowski and Unruh modes, the latter being the eigenmodes of the Rindler 
energy in each Rindler wedge, for the electromagnetic (EM) field.  
One can then deduce how the Unruh creation operators are expressed in terms of the Minkowski ones. This result will be used in the next section to calculate the emitted power in the Rindler frame. Here we focus on setting up the problem and presenting the main relations. We leave all the technical details to the Appendices.

In Rindler coordinates, the Minkowski line element takes the form \cite{Bire82}:
\begin{equation}
    ds^2 = e^{2 a \xi} (d \tau^2 - d \xi^2) - dx^2 - dy^2\,,
\label{metric}    
\end{equation}
where the coordinates $\tau$ and $\xi$ are defined by $t = a^{-1} e^{a \xi} \sinh{a \tau}$ and $z = a^{-1} e^{a \xi} \cosh{a \tau}$ with $a>0$. The part of Minkowski spacetime covered by the metric \eqref{metric} is restricted by $z > |t|$ and is known as the right Rindler wedge. The proper acceleration of the world lines with $\xi$, $x$ and $y$ constant is given by $a e^{-a \xi}$ and therefore uniformly accelerated observers follow these world lines. Similarly, the Rindler coordinates $(\overline{\tau},\overline{\xi})$ which cover the part of Minkowski spacetime with $z < -|t|$, known as the left Rindler wedge, are defined by $t = a^{-1} e^{a \overline{\xi}} \sinh{a \overline{\tau}}$ and $z = -a^{-1} e^{a \overline{\xi}} \cosh{a \overline{\tau}}$.

The Lagrangian describing the EM field in the Feynman gauge is
\begin{equation}
    \mathcal{L} = -\frac{1}{4} \sqrt{-g} F_{\mu \nu}F^{\mu \nu} -\frac{1}{2} \sqrt{-g}(\nabla_{\alpha} A^{\alpha})^2\,,
\label{lagrangian}    
\end{equation}
where $F_{\mu \nu} = \nabla_{\mu} A_{\nu}- \nabla_{\nu} A_{\mu}$ and the second term is a gauge fixing term. For the metric \eqref{metric}, $\sqrt{-g} = e^{2 a \xi}$. The equations of motion of the Lagrangian \eqref{lagrangian} are given by
\begin{equation}
    \nabla^{\mu} (\nabla_{\mu} A_\rho) - R_{\rho}^{\; \; \lambda} A_{\lambda} = 0\,, 
\label{general-equation-of-motion}
\end{equation}
where $R_{\mu \nu}$ is the Ricci tensor.
Note that the Lagrangian (\ref{lagrangian}) does not include any mutual interaction between the 
point particle and the quantum field. The back-reaction effects arising from such 
a term are important in determining radiation reaction contributions in the inertial frame \cite{Brodin:2007vf,Lin.2005,Lin.2007,Lynch:2019hmk}, but are neglected in our analysis as subdominant. We have $R_{\mu\nu}=0$ in Eq.~\eqref{general-equation-of-motion} since Minkowski spacetime is flat.  Thus, the equations of motion simplify to 
\begin{equation}
    \nabla^{\mu} \nabla_{\mu} A_\rho = 0 \, .
    \label{eom}
\end{equation}
Expanding the EM field operator in the right Rindler wedge gives
\begin{equation}
    \hat{A}^{R}_{\mu} (x^\nu) = \int d^2 \mathbf{k}_\perp  \; d \omega \sum_{\lambda = 1}^4 \left[ a^R_{(\lambda, \omega, \mathbf{k}_\perp)} A^{R(\lambda, \omega, \mathbf{k}_\perp)}_{\mu}(x^\nu) + \rm{h.c.}\right]\,,
\end{equation}
where $\mathbf{k}_\perp = (k_x,k_y) \neq (0,0)$, $\omega >0$, and where $a^R_{(\lambda, \omega, \mathbf{k}_\perp)}$ is the annihilation operator in the right Rindler wedge and the index $\lambda$ labels the different polarizations. 
The modes $A^{R(\lambda, \omega, \mathbf{k}_\perp)}_{\mu}$ solve Eq.~\eqref{eom} and are given in Refs.~\cite{PhysRevD.45.R3308, Higuchi:1992td}, with the notation
$A_\mu = (A_\tau,A_\xi,A_x,A_y)$, by
\begin{equation}
\begin{aligned}
A^{R(\text{I}, \omega,\mathbf{k}_\perp)}_{\mu} & = k_\perp^{-1} (0,0,k_y v^R_{\omega \mathbf{k}_\perp}, - k_x v^R_{\omega \mathbf{k}_\perp})\, , \\ 
A^{R(\text{II}, \omega, \mathbf{k}_\perp)}_{\mu} & = k_\perp^{-1} (\partial_{\xi} v^R_{\omega \mathbf{k}_\perp},\partial_\tau v^R_{\omega \mathbf{k}_\perp},0,0)\,, \\
A^{R(G, \omega, \mathbf{k}_\perp)}_{\mu} & = k_\perp^{-1} \nabla_\mu v^R_{\omega \mathbf{k}_\perp}, \\
A^{R(L, \omega,\mathbf{k}_\perp)}_{\mu} & = k_\perp^{-1} (0,0,k_x v^R_{\omega \mathbf{k}_\perp}, k_y v^R_{\omega \mathbf{k}_\perp})\,, \\
\end{aligned}
\label{modesolution}
\end{equation}
where $k_\perp = \sqrt{k_x^2+k_y^2}$ is the transverse momentum and $v^R_{\omega \mathbf{k}_\perp}$ is the solution to the scalar Klein-Gordon equation $\square \phi = 0$ and is given by
\begin{equation}
v^R_{\omega \mathbf{k}_\perp} = \sqrt{\frac{\sinh(\pi \omega/a)}{4 \pi^4 a}} K_{i \omega/a}\left( \frac{k_\perp e^{a\xi}}{a}\right) e^{- i \omega \tau + i \mathbf{k}_\perp \cdot \mathbf{x}_\perp}\,.
\label{scalarsolution}
\end{equation}
Here, the function $K_\nu(z)$ is the modified Bessel function of the second kind.
The vacuum annihilated by $a^R_{(\lambda, \omega, \mathbf{k}_\perp)}$ is referred to as the Rindler, or Fulling, vacuum and is denoted by $\ket{0_{\text{R}}}$. 
It differs from the Minkowski vacuum $\ket{0_{\text{M}}}$ since the Rindler modes $A^{R(\lambda, \omega, \mathbf{k}_\perp)}_{\mu}$ are not a combination of purely 
positive-frequency Minkowski modes but contain negative-frequency modes as well~\cite{Bire82}.

The normalization of the physical modes 
$A^{R(\text{I}, \omega, \mathbf{k}_\perp)}_{\mu}$ and $A^{R(\text{II}, \omega, \mathbf{k}_\perp)}_{\mu}$, which satisfy the
Lorenz condition $\nabla_\mu A^\mu = 0$ and are not pure gauge, is 
determined with respect to the Klein-Gordon inner product (see Appendix~\ref{app:A}). As a result~\cite{PhysRevD.45.R3308,Higuchi:1992td} the creation and annihilation operators for the physical modes satisify the following commutation relations:
\begin{equation}
\left[ a^R_{(\lambda,\omega,\mathbf{k}_\perp)},a^{R\dagger}_{(\lambda',\omega',\mathbf{k}_\perp')}\right] = \delta_{\lambda\lambda'}\delta(\omega-\omega')\delta^{(2)}(\mathbf{k}_\perp-\mathbf{k}'_\perp)\,.
\end{equation}

As we shall see later, the only non-zero components of the current $j^\mu$ representing a charge uniformly accelerated in the $z$-direction are $j^\tau$ and $j^\xi$ [see Eq.~\eqref{current}]. This implies that the only modes that couple to this current
are the second physical modes $A^{R(\text{II}, \omega, \mathbf{k}_\perp)}_{\mu}$: 
the $\tau$- and $\xi$-components of the 
modes $A^{R(\text{I}, \omega,\mathbf{k}_\perp)}_{\mu}$ and $A^{R(L, \omega,\mathbf{k}_\perp)}_{\mu}$ are zero, and the coupling of $j^\mu$ to the modes
$A^{R(G, \omega, \mathbf{k}_\perp)}_{\mu}$ vanishes because of the conservation equation $\nabla_\mu j^\mu = 0$.  Thus, we need to consider only the modes
$A^{R(\text{II}, \omega, \mathbf{k}_\perp)}_{\mu}$.
They can be written as  $A_\mu^{R(\text{II},\omega, \mathbf{k}_\perp)} = k_\perp^{-1} \epsilon_{\mu\nu} \partial^\nu v^R_{\omega \mathbf{k}_\perp}$, where 
$\epsilon_{\mu \nu}$ is the anti-symmetric tensor on the plane in Minkowski spacetime with $x$ and $y$ fixed, which has the following metric:
\begin{equation}
    ds_{(2)}^2 = dt^2 - dz^2 = e^{2a\xi}(d\tau^2 - d\xi^2)\,,
\end{equation}
with $\epsilon_{zt}=1$.
Therefore, in Minkowski coordinates these modes take the following form:
\begin{equation}
 A_\mu^{R(\text{II},\omega, \mathbf{k}_\perp)} = k_\perp^{-1} ( \partial_zv^R_{\omega \mathbf{k}_\perp},\partial_tv^R_{\omega \mathbf{k}_\perp},0,0)\,, 
 \label{rightmincoor}
\end{equation} 
in the notation $A_\mu = (A_t,A_z,A_x,A_y)$. Using a similar approach on the left Rindler wedge (see Appendix~\ref{app:B}), the left EM modes $A_\mu^{L(\lambda,\omega, \mathbf{k}_\perp)} $can be obtained from the right ones with the substitution $v^R_{\omega \mathbf{k}_\perp} \to v^L_{\omega \mathbf{k}_\perp}$ where $v^L_{\omega \mathbf{k}_\perp}$ are the corresponding solutions to the scalar Klein-Gordon equation in the left Rindler wedge. The purely positive-frequency EM modes, or Unruh modes are then \cite{PhysRevD.106.065002}:
\begin{equation}
    \begin{aligned}
        W_\mu^{(\lambda,-,\omega, \mathbf{k}_\perp)} = \frac{A_\mu^{R(\lambda,\omega, \mathbf{k}_\perp)} + e^{- \pi \omega/a} A_\mu^{L(\lambda,\omega, -\mathbf{k}_\perp)*}}{\sqrt{1 - e^{-2\pi \omega/a}}}\,, \\
        W_\mu^{(\lambda,+,\omega, \mathbf{k}_\perp)} = \frac{A_\mu^{L(\lambda,\omega, \mathbf{k}_\perp)} + e^{- \pi \omega/a} A_\mu^{R(\lambda,\omega, -\mathbf{k}_\perp)*}}{\sqrt{1 - e^{-2\pi \omega/a}}}\,.
    \end{aligned}
    \label{Unruhmodes}
\end{equation}
The full EM field can be expanded in terms of these modes as they form a complete set:
\begin{equation}
\begin{aligned}
    \hat{A}_\mu = \int d^2 \mathbf{k}_\perp \int_0^{+\infty} d \omega \sum_\lambda \left[ W^{(\lambda, -,\omega, \mathbf{k}_\perp)*}_\mu a^\dagger_{(\lambda, -,\omega, \mathbf{k}_\perp)} \right. \\ 
    \left. +  W^{(\lambda,+, \omega, \mathbf{k}_\perp)*}_\mu a^\dagger_{(\lambda, +,\omega, \mathbf{k}_\perp)}+ \rm{h.c.} \right]\,,
    \label{emexpansionW}
\end{aligned}
\end{equation}
where $a^\dagger_{(\lambda, \pm,\omega, \mathbf{k}_\perp)}$ are the Unruh creation operators. The creation operators for the second physical modes can be expanded in terms of the Minkowski ones $b^{\dagger}_{\mathbf{k}}$ with the
momentum $\mathbf{k}$ and the  polarization vector $\varepsilon^{\mu}(\mathbf{k}) = k_\perp^{-1}(k_z,k_0,0,0)$ that satisfy $[b_{\mathbf{k}},b^\dagger_{\mathbf{k}'}] =\delta^{(3)}(\mathbf{k}-\mathbf{k}')$ as (see Appendix~\ref{app:B}) 
\begin{eqnarray}
a^\dagger_{(\text{II}, \pm,\omega, \mathbf{k}_\perp)} =  i\int_{-\infty}^{+\infty}  \frac{d k_z}{\sqrt{2 \pi a k_0}}  e^{\pm i \vartheta(k_z)\omega} b^{\dagger}_{\mathbf{k}}.
\label{operatorexpansion}
\end{eqnarray}
where $\vartheta(k_z) = (2a)^{-1} \ln \{(k_0+k_z)/(k_0-k_z)\}$ is the normalized rapidity and $k_0 = \sqrt{k_\perp^2 + k_z^2}$ is the total energy of the photon.
 
\section{Photon emission in the Rindler frame}
\label{phoemi}

The interaction between a photon and the charged particle (an electron for example) in the right Rindler wedge with the associated 
classical current $j^\mu$ is described by the action:
\begin{equation}
    S_I = -\int d^4 x \sqrt{-g} j^{\mu} \hat{A}^R_{\mu}\,.
\label{interaac}    
\end{equation}
We consider a point charge $q$ located in the right Rindler wedge at $\xi = x =y =0$.  This charge is uniformly accelerated with proper acceleration $a$.  It would be ideal to
consider a charge uniformly accelerated only for a finite time.  However, such a charge would enter the right Rindler wedge at $\tau=-\infty$ and leave it at 
$\tau=+\infty$.  This behavior of the charge would make the analysis rather involved.  Instead, we consider a point charge which is charged and uncharged 
through a wire at $x=y=0$
extending from $\xi=0$ to $+\infty$ (see Refs.~\cite{PhysRevD.45.R3308,Higuchi:1992td} for a similar model).
The associated current $j^{\mu}$ is:
\begin{equation}
    \begin{aligned}
        j^{\tau} & = q F(\tau) \delta(\xi) \delta(x)\delta(y)\,, \\
        j^{\xi} & = -q F'(\tau) e^{-2a\xi}\theta(\xi) \delta(x)\delta(y)\,, \\
        j^x  & = j^y =0\,,
    \end{aligned}
    \label{current}
\end{equation}
where $F(\tau)$ is a smooth function.  A charge uniformly accelerated forever 
corresponds to the choice $F(\tau) =1$ for all $\tau$.  This choice would lead to inconsistencies
even for classical Larmor radiation \cite{Ritus,Schwinger}. Considering a finite period of acceleration with non-zero charge avoids these inconsistencies.
Thus, the function $F(\tau)$ is chosen in such a way to ensure that the period where the particle has non-zero charge is finite. We let $F(\tau)=1$ 
for $|\tau|< T $, where $2T \gg 1/a$ is the period of constant charge. For $|\tau| > T + b$, where $1/a \ll b\ll T$, we let the particle have no charge, i.e.\ $F(\tau)=0$.  
The period 
$T< |\tau|< T+ b$ is a period of smooth transition between the two. In the end, we let $T \to +\infty$ but keep $b$ finite, [see Eq.~\eqref{uni-acc-rate}], thus removing contributions to the transition rate (with fixed transverse momentum) coming from transition effects. 
Note that the current~\eqref{current} satisfies current conservation $\nabla_\mu j^\mu =0 $, which ensures gauge invariance. 

At tree level,  the amplitude of emission of a photon in the Rindler vacuum state by the charged particle is given by:
\begin{equation}
    \mathcal{A}^{e}_{( \omega, \mathbf{k}_\perp)} = i \bra{\text{II}, \omega, \mathbf{k}_\perp}  S_I \ket{0_\text{R}}\,,
\label{ampli1}    
\end{equation}
where $\ket{\text{II}, \omega, \mathbf{k}_\perp} = a^{R \dagger}_{(\text{II},\omega,\mathbf{k}_\perp)} \ket{0_{\text{R}}}$. The emission amplitude can be computed explicitly combining Eqs.~\eqref{modesolution}, \eqref{scalarsolution} and \eqref{current}:
\begin{equation}
\begin{aligned}
\mathcal{A}^{e}_{( \omega, \mathbf{k}_\perp)} = -i q \Tilde{F}(\omega)\sqrt{\frac{\sinh(\pi\omega/a)}{4\pi^4a}}
\left[ K'_{i \omega/a} \left( \frac{k_\perp}{a}\right) \right.\\
-\left. \frac{\omega^2}{k_\perp} \int_0^{+\infty} d \xi K_{i \omega/a} \left( \frac{k_\perp e^{a\xi}}{a}\right)\right]\,,
\end{aligned}
\label{finalemissionampli}
\end{equation}
where $\Tilde{F}$, the Fourier transform of $F$, is defined by
\begin{equation}
    \Tilde{F}(\omega) = \int_{- \infty}^{+\infty} d \tau F(\tau) e^{i \omega \tau}\,.
    \label{fourier}
\end{equation}
The amplitude for the absorption of a photon with transverse momentum $- \mathbf{k}_\perp$ is
\begin{equation}
     \mathcal{A}^{a}_{( \omega, -\mathbf{k}_\perp)} = i \bra{0_R} S_I \ket{\text{II}, \omega, -\mathbf{k}_\perp}\,.
\end{equation}

The total one-photon interaction probability is found by taking into account the Unruh effect, i.e., the fact that, in the Rindler wedge, 
the Minkowski vacuum state
is equivalent to a thermal bath of temperature $a/2\pi$ with the Bose-Einstein distribution function $(e^{2\pi\omega/a}- 1)^{-1}$ with respect to the
Rindler energy.
The result is
\begin{equation}
    \begin{aligned}
        P_{\text{tot}} & = \int_0^{+\infty}d\omega \int d^2\mathbf{k}_\perp
         \left[\frac{\left|\mathcal{A}^e_{(\omega,\mathbf{k}_\perp)}\right|^2}{1- e^{-2\pi\omega/a}} + 
        \frac{\left|\mathcal{A}^a_{(\omega,-\mathbf{k}_\perp)}\right|^2}{e^{2\pi\omega/a} - 1}\right]\,.
    \end{aligned}
    \label{prob}
\end{equation}
Note that
\begin{equation}
    \frac{1}{1-e^{-2\pi\omega/a}} = 1 + \frac{1}{e^{2\pi\omega/a} -1}\,.
\end{equation}
Thus, the first term in the integrand gives the (spontaneous and induced) photon emission probability 
while the second term gives the photon absorption probability in the presence of
the FDU thermal bath of temperature $a/2\pi$.

To understand the Larmor formula in the context of the Unruh effect, we first note that one can interpret 
Eq.~\eqref{prob} as the norm squared of a one-photon final state expressed as a linear combination of Unruh states 
$a_{(\mathrm{II},\pm,\omega,\mathbf{k}_\perp)}^\dagger\ket{0_{\mathrm{M}}}$ as shown in Appendix~\ref{app:C}:
\begin{widetext}
\begin{equation}
\begin{aligned}
   \ket{f_{1\textrm{-photon}}}  = \int d^2\mathbf{k}_\perp\int_{0}^{+\infty}d\omega
   \left[\frac{\mathcal{A}^e_{(\omega,\mathbf{k}_\perp)}a_{(\mathrm{II},-,\omega,\mathbf{k}_\perp)}^\dagger}{\sqrt{1-e^{-2\pi\omega/a}}}
+ \frac{\mathcal{A}^a_{(\omega,-\mathbf{k}_\perp)}a_{(\mathrm{II}, +,\omega,\mathbf{k}_\perp)}^\dagger}{\sqrt{e^{2\pi\omega/a}-1}}\right]\ket{0_{\mathrm{M}}}\,.
\label{main:1-photon-Rindler}
\end{aligned}
\end{equation}
\end{widetext}
This state can be expressed as a linear combination of states $b_{\mathbf{k}}^\dagger\ket{0_\mathrm{M}}$ with (Minkowski) momentum $\mathbf{k}$ using
Eq.~\eqref{operatorexpansion} as
\begin{equation}
\begin{aligned}
   \ket{f_{1\textrm{-photon}}} & = i\int d^2\mathbf{k}_\perp
    \int_{-\infty}^{+\infty} \frac{dk_z}{\sqrt{2\pi ak_0}}   \\
    & \quad \times \int_{-\infty}^{+\infty}d\omega
    \frac{e^{-i\vartheta(k_z)\omega}\mathcal{A}^e_{(\omega,\mathbf{k}_\perp)}}{\sqrt{1-e^{-2\pi\omega/a}}}b_\mathbf{k}^\dagger\ket{0_{\text{M}}}\,,
    \label{main:1-photon-Minkowski}
\end{aligned}
\end{equation}
where we have used the relation
\begin{equation} 
    \frac{\mathcal{A}^a_{(\omega,-\mathbf{k}_\perp)}}{{\sqrt{e^{2\pi\omega/a} - 1}}}
= \frac{\mathcal{A}^e_{(-\omega,\mathbf{k}_\perp)}}{\sqrt{1-e^{2\pi\omega/a}}}\,.
\label{relAatoAe}
\end{equation}
Then, one can write Eq.~\eqref{prob} as
\begin{equation}
\begin{aligned}
 P_{\text{tot}}   &  =   \langle f_{1\textrm{-photon}}|f_{1\textrm{-photon}}\rangle \\
 & = \int d^2\mathbf{k}_\perp
    \int_{-\infty}^{+\infty} \frac{dk_z}{2\pi ak_0} 
 \left|\int_{-\infty}^{+\infty}d\omega\frac{e^{-i\vartheta(k_z)\omega}\mathcal{A}^e_{(\omega,\mathbf{k}_\perp)}}{\sqrt{1-e^{-2\pi\omega/a}}}\right|^2.
\end{aligned}
\label{firstexpproba}
\end{equation}
Notice that there are interference terms between the emission and absorption in the co-accelerated frame.
Substituting Eq.~\eqref{finalemissionampli} into this equation, we find
\begin{equation}
    P_{\text{tot}} = \frac{a}{16\pi^3}\int d^2\mathbf{k}_\perp \int_{-\infty}^{+\infty}d\vartheta\,\left|\mathcal{A}(\mathbf{k})\right|^2\,,
    \label{total-em-prob}
\end{equation}
where we used $d\vartheta = dk_z/ak_0$ and where
\begin{equation}
\begin{aligned}
    \mathcal{A}(\mathbf{k}) & = -\frac{q}{\pi a}\int_{- \infty}^{+\infty} d \omega \Tilde{F}(\omega)e^{- i \omega \vartheta} e^{\pi \omega/2a} \\
    &  \times \left[ K'_{i \omega/a} \left( \frac{k_\perp}{a}\right) - \frac{\omega^2}{k_\perp} \int_0^{+\infty} d \xi K_{i \omega/a} \left( \frac{k_\perp e^{a\xi}}{a}\right)\right]\,.
\end{aligned}
\label{amplitude-for-k}
\end{equation}
To identify the contribution to the amplitude $\mathcal{A}(\mathbf{k})$ from the period of uniform acceleration, separating out the contribution
from the transition period, we need to express this amplitude in terms of $F(\tau)$ rather than $\Tilde{F}(\omega)$. The result is (see Appendix~\ref{app:D})
\begin{equation}
\begin{aligned}
    \mathcal{A}(\mathbf{k}) & = \frac{qa}{k_\perp}\int_{-\infty}^{+\infty} d\tau
    \left\{ \frac{F(\tau)e^{-i(k_\perp/a)\sinh a(\vartheta-\tau)}}{\cosh^2 a(\vartheta-\tau)} \right. \\    
   &  \quad \left. - \frac{i}{a^3} \frac{d}{d \tau} \left[ \frac{1}{\cosh a(\vartheta-\tau)} \frac{d}{d \tau} \left( \frac{F'(\tau)}{\cosh^2 a(\vartheta-\tau)}\right)\right]\right.\\
   & \left.\quad\quad \quad \times \int_{k_\perp/a}^{+\infty}\frac{dz}{z^2}e^{-iz\sinh a(\vartheta-\tau)}\right\}\,.
\end{aligned}
\label{proba3terms}
\end{equation}
Due to our choice of the function $F(\tau)$ stated before and by further letting $(-c_1,-c_2)\cup (c_1,c_2) = \{ \tau\in\mathbb{R}: 0 < F(\tau) <1\}$ be such that $c_1 -T, T+b-c_2 \gg 1/a$, we can conclude that
\begin{equation}
\begin{aligned}
    \mathcal{A}(\mathbf{k}) & \approx \frac{qa}{k_\perp}\int_{-\infty}^{+\infty} d\tau\,\frac{F(\tau)}{\cosh^2 a(\vartheta-\tau)}e^{-i(k_\perp/a)\sinh a(\vartheta-\tau)}\\
    & \hspace{5cm} \text{if}\ |\vartheta| < T\,,
\end{aligned}
\label{agrees-with-standard}
\end{equation}
and $\mathcal{A}(\mathbf{k})\approx 0$ if $|\vartheta| > T+b$, for each $\mathbf{k}$ with $k_\perp >0$, because of the exponential decay of each term in Eq.~\eqref{proba3terms}.   The amplitude $\mathcal{A}(\mathbf{k})$ is a continuous function of $\vartheta$.  Hence, the $\vartheta$-integral of 
$|\mathcal{A}(\mathbf{k})|^2$ for $T < |\vartheta| < T+b$ is finite.  Furthermore, if we let $T\to +\infty$ while keeping the shape
of the function $F(\tau)$ in the transition period unchanged, then the $\vartheta$-integral of $|\mathcal{A}(\mathbf{k})|^2$ 
over this period will remain constant.

We define the emission probability with $\mathbf{k}_\perp$ fixed by
\begin{equation}\label{eq:emission-probability}
    P(k_\perp) = \frac{a}{16\pi^3}
    \int_{-\infty}^{+\infty}d\vartheta\,\left|\mathcal{A}(\mathbf{k})\right|^2\,.
\end{equation}
    Then, the emission rate with $\mathbf{k}_\perp\,(\neq\mathbf{0})$ fixed is~\footnote{The total emission rate obtained by integrating $R(k_\perp)$ over
$\mathbf{k}_\perp$ diverges due to the contribution from small $k_\perp$. The total energy emitted, which is
obtained by multiplying the integrand of Eq.~\eqref{total-em-prob} by $k_0$, also appears to be infinite due to the contribution from the second term
in Eq.~\eqref{proba3terms}.  One way to make sure that the total emitted energy be 
finite is to introduce a negative charge at $\xi=L < \infty$ and
let the current flow only between the two charges at $\xi=0$ and $L$.  However, the total energy emitted during the period of uniform
acceleration is finite and agrees with the Larmor formula after the initial and final effects are removed as we shall see.}
\begin{equation}
\begin{aligned}
    R(k_\perp) & = \lim_{T\to +\infty}\frac{P(k_\perp)}{2T} \\
    & = \lim_{T\to +\infty}\frac{1}{2T}\times\frac{q^2 a^3}{16\pi^3 k_\perp^2}\int_{-\infty}^{+\infty} d\vartheta \\
& \quad \times    \left| \int_{-\infty}^{+\infty} d\tau\,\frac{F(\tau)e^{-i(k_\perp/a)\sinh a(\vartheta-\tau)}}{\cosh^2 a(\vartheta-\tau)}\right|^2\,.
\end{aligned}
\label{uni-acc-rate}
\end{equation}
 We note in passing that Eq.~\eqref{agrees-with-standard} with $F(\tau)=1$ 
 agrees with the amplitude for the general motion which can be straightforwardly derived from Ref.~\cite[Eq.~(2.33)]{Higuchi:2009ms} and is given by
\begin{equation}
 \mathcal{A}^\mu(\mathbf{k}) = - q\int_{-\infty}^{+\infty} \frac{d\tau}{k\cdot v}\left( a^\mu - \frac{k\cdot a}{k\cdot v}v^\mu\right)e^{ik\cdot x}\,,
\end{equation}
where $x^\mu(\tau)$, $v^\mu(\tau)$ and $a^\mu(\tau)$ are the world line of the charge, its $4$-velocity and $4$-acceleration, respectively,
with the identification $\mathcal{A}^\mu(\mathbf{k}) = -\mathcal{A}(\mathbf{k})\varepsilon^\mu(\mathbf{k})$. The emission rate is given, in the large $T$ limit, (see Appendix~\ref{app:E}) by
\begin{equation}
\begin{aligned}
    R(k_\perp) & = \frac{q^2 a^3}{16\pi^3 k_\perp^2}\int_{-\infty}^{+\infty} d\overline{\vartheta}\int_{- \infty}^{+\infty} d \sigma\\
& \quad \times 
\frac{e^{ 2i (k_\perp/a)\cosh{a\overline{\vartheta}}\sinh{a\sigma/2}}}{[\cosh^2{a\overline{\vartheta}} + \sinh^2{a \sigma/2}]^2}\,,  
\end{aligned}
\label{R-per-k-perp}
\end{equation}
where $\Bar{\vartheta}$ is the rapidity in the rest frame of the charge. First we verify that Eq.~\eqref{R-per-k-perp} agrees with the result of Ref.~\cite{PhysRevD.45.R3308,Higuchi:1992td} by the change of
integration variables $s_{\pm}=\overline{\vartheta}\pm \sigma/2$, which essentially restores the original 
expression~\eqref{uni-acc-rate}.  Thus, we find
\begin{equation}
    \begin{aligned}
        R(k_\perp) & = \frac{q^2 a^3}{16\pi^3 k_\perp^2}\left| \int_{-\infty}^{+\infty}\frac{e^{i(k_\perp/a)\sinh as}}{\cosh^2 as}ds\right|^2 \\
        & = \frac{q^2}{4\pi^3 a}\left|K_1\left(\frac{k_\perp}{a}\right)\right|^2\,,
    \end{aligned}
    \label{eq:Rk-perp}
\end{equation}
as expected.  The second equality can be established using Ref.~\cite[Eq.~8.432.5]{gradshteyn2014table}.

The power radiated in the rest frame of the charge is given by multiplying the integrand of Eq.~\eqref{R-per-k-perp} by
$\overline{k}_0 = k_\perp\cosh a\overline{\vartheta}$ and integrating the result over $\mathbf{k}_\perp$.  Thus,
defining the energy and longitudinal momentum in the rest frame by $\overline{k}_0 = k_\perp\cosh a\overline{\vartheta}$ and
$\overline{k}_z = k_\perp\sinh a\overline{\vartheta}$, respectively, we find 
\begin{equation}
    \begin{aligned}
        S_{\text{rest}} & = \frac{q^2 a^2}{16\pi^3}\int d^2\mathbf{k}_\perp d\overline{k}_z\,k_\perp^2
        \int_{- \infty}^{+\infty} d \sigma\\
& \quad \times 
\frac{\cos[2(\overline{k}_0/a)\sinh(a\sigma/2)]}{[\overline{k}_0^2 + k_\perp^2\sinh^2(a \sigma/2)]^2}\,.  
    \end{aligned}
\end{equation}
Then, by writing 
$d^2\mathbf{k}_\perp d\overline{k}_z = d\overline{k}_0 \overline{k}_0^2 d\Omega$, where $d\Omega$ is the solid angle element
in the instantaneous rest frame of the accelerated particle, and  where $k_\perp = \overline{k}_0 \sin\theta$, we find
\begin{align}
        \frac{dS_\text{rest}}{d\Omega} & = \frac{q^2 a^2}{32\pi^3}\sin^2\theta \notag \\
        & \quad \times \int_{-\infty}^{+\infty} d\sigma 
        \int_{-\infty}^{+\infty} d\overline{k}_0 \frac{e^{2i(\overline{k}_0/a)\sinh (a\sigma/2)}}{[1 + \sin^2\theta\sinh^2 (a\sigma/2)]^2} 
        \notag\\
        & = \frac{q^2 a^2}{16\pi^2}\sin^2\theta.
\end{align}
This is the well-known Larmor formula with
\begin{equation}
    S_{\text{rest}} = \frac{q^2 a^2}{6\pi}\,.
\end{equation}

\section{Conclusions}

In this paper, we studied the electromagnetic radiation from a uniformly accelerated charge, the Larmor radiation, 
in the context of the Unruh effect, i.e.\ the fact that the Minkowski vacuum state appears to be a thermal bath to a uniformly 
accelerate observer.  A formal derivation of the power radiated from a charge uniformly accelerated forever
does not lead to the correct Larmor formula.  For this reason we studied a model where a non-zero charge is accelerated only for
a finite time and identified the part of the radiation due to the period in which the non-zero charge has a uniform acceleration, removing the transition effects at the start and the end. Then we took the infinite-time limit to recover the Larmor formula.

We used the observation of Unruh and Wald~\cite{PhysRevD.29.1047} that both the emission and absorption of a photon in the Rindler
frame correspond to emission of a photon in the inertial frame.  Thus, a uniformly accelerated charge emit a photon in the Unruh
modes, which can be decomposed into the usual Minkowski modes with definite momenta. In this manner we were able to reproduce the
Larmor radiation formula for the power emitted from a uniformly accelerated charged particle.

Larmor's formula was found previously
in Refs.~\cite{Lynch:2019hmk,PhysRevD.106.065002} for photons in the laboratory frame and in
Refs.~\cite{Ren1994,PhysRevD.102.105016}, for scalar fields in the context of the Unruh effect. Our derivation
makes the link between the Unruh effect and the Larmor radiation from a uniformly accelerated charged particle clearer and will help
in resolving some of the controversies that have surrounded the Unruh effect since its discovery. 
  
\begin{acknowledgments}
We thank Prof.~Antonino Di Piazza and Prof.~Subir Sarkar for the helpful discussions and comments.  This work was supported in part by EPSRC grants no. EP/X01133X/1 and EP/X010791/1. GG is also a member of the Quantum Sensing for the Hidden Sector (QSHS) collaboration, supported by STFC grant no. ST/T006277/1.
\end{acknowledgments}

\appendix

\section{Normalization of the right Rindler modes }
\label{app:A}

The normalization of the physical modes 
$A^{R(\text{I}, \omega, \mathbf{k}_\perp)}_{\mu}$ and $A^{R(\text{II}, \omega, \mathbf{k}_\perp)}_{\mu}$, which satisfy the
Lorenz condition $\nabla_\mu A^\mu = 0$ and are not pure gauge, is 
determined with respect to the Klein-Gordon inner product:
\begin{equation}
\left(A^{R(i)},A^{R(j)}\right) = \int_{\Sigma} d \Sigma_{\mu} \; \Xi^{\mu} [A^{R(i)},A^{R(j)} ]\,,
\label{KG-innner-product_app}
\end{equation}
where the labels $i,j$ represent $(\lambda, \omega, \mathbf{k}_\perp)$, and $\Sigma$ is a Cauchy hypersurface $\tau = \emph{constant}$.
The vector $\Xi^{\mu} [A^{R(i)},A^{R(j)} ]$ is given by
\begin{equation}
\Xi^{\mu} [A^{R(i)},A^{R(j)} ] = \frac{i}{\sqrt{-g}} \left( A^{R(i)*}_\nu \pi^{R(j) \mu \nu} - A^{R(j)}_\nu \pi^{R(i) *\mu \nu}  \right)\,,    
\end{equation}
where $\pi^{R(i) \mu \nu} =  \partial \mathcal{L} / \partial_\mu A_\nu |_{A^{R(i)}_\mu}$ and the asterisk indicates complex conjugation. 
This vector satisfies the conservation equation $\nabla_\mu\Xi^{\mu} [A^{R(i)},A^{R(j)} ]=0 $ and, hence, the Klein-Gordon inner product~\eqref{KG-innner-product_app}
is $\tau$-independent. The normalization of the physical modes  $A^{R(\text{I}, \omega, \mathbf{k}_\perp)}_{\mu}$ and $A^{R(\text{II}, \omega, \mathbf{k}_\perp)}_{\mu}$
are chosen such that
\begin{equation}
    \left(A^{R(\lambda,\omega,\mathbf{k}_\perp)},A^{R(\lambda',\omega',\mathbf{k}_\perp')}\right) = \delta_{\lambda\lambda'}\delta(\omega-\omega')\delta^{(2)}(\mathbf{k}_\perp-\mathbf{k}'_\perp)\,.
\end{equation} 

\section{Unruh and Minkowski creation operators}
\label{app:B}

In this Appendix we derive the relation between the Unruh and Minkowksi creation operators. Recall that the right second physical modes $\lambda = \text{II}$ can be written as 
\begin{equation}
 A_\mu^{R(\text{II},\omega, \mathbf{k}_\perp)} = k_\perp^{-1} ( \partial_zv^R_{\omega \mathbf{k}_\perp},\partial_tv^R_{\omega \mathbf{k}_\perp},0,0)\,. 
 \label{rightmincoor_app}
\end{equation}
The left scalar modes $v^L_{\omega \mathbf{k}_\perp}$, which are non-zero in the left Rindler wedge and vanish in the right one, are obtained from $v^R_{\omega \mathbf{k}_\perp}$ by letting $z \to -z$. (The right Rindler modes $v^R_{\omega\mathbf{k}_\perp}$ vanish in the left Rindler wedge by definition.)  They can be found by replacing $\tau$ by $\overline{\tau}$ and $\xi$ by $\overline{\xi}$ in the expression of $v^R_{\omega \mathbf{k}_\perp}$, where $\overline{\tau}$ and $\overline{\xi}$ are the left Rindler coordinates. Then the left Rindler EM modes can be obtained from the right ones by simply replacing $v^R_{\omega \mathbf{k}_\perp}$ by $v^L_{\omega \mathbf{k}_\perp}$. In particular, the second physical left Rindler modes are given by: 
\begin{equation}
 A_\mu^{L(\text{II},\omega, \mathbf{k}_\perp)} = k_\perp^{-1} ( \partial_zv^L_{\omega \mathbf{k}_\perp},\partial_tv^L_{\omega \mathbf{k}_\perp},0,0)\,. 
 \label{leftmincoor_app}
\end{equation}
Similarly to the right Rindler modes, the left Rindler modes are not purely positive-frequency with respect to the inertial time $t$. However, in the scalar case, the purely positive-frequency modes, or Unruh modes, are linear combinations of left and right Rindler modes and are given by:
\begin{equation}
    \begin{aligned}
        w_{- \omega \mathbf{k}_\perp} & = \frac{v^R_{\omega \mathbf{k}_\perp} + e^{- \pi \omega/a} v^{L*}_{\omega -\mathbf{k}_\perp}}{\sqrt{1 - e^{-2\pi \omega/a}}}\,, \\
        w_{+ \omega \mathbf{k}_\perp} & = \frac{v^L_{\omega \mathbf{k}_\perp} + e^{- \pi \omega/a} v^{R*}_{\omega -\mathbf{k}_\perp}}{\sqrt{1 - e^{-2\pi \omega/a}}}\,.
    \end{aligned}
    \label{scalarunruhsuppmat_app}
\end{equation}
They form a complete set of orthonormal solutions to the scalar Klein-Gordon equation. The second physical EM Unruh mode, $\lambda = \text{II}$, can be found by combining Eqs.~\eqref{rightmincoor_app}, \eqref{leftmincoor_app} and \eqref{scalarunruhsuppmat_app}. It
is given in terms of the scalar Unruh modes as:
\begin{equation}
 W_\mu^{(\text{II},\pm, \omega, \mathbf{k}_\perp)} = k_\perp^{-1} ( \partial_z w_{\pm\omega \mathbf{k}_\perp},\partial_tw_{\pm\omega \mathbf{k}_\perp},0,0)\,. 
 \label{unruhemscalar_app}
\end{equation}

To find the relation between the Unruh and the Minkowski creation operators, we need to find the relation between the Unruh and the Minkowski modes. For this purpose, we make use of the expansion of  the scalar positive-frequency modes~\cite{Crispino2008}:
\begin{equation}
 w_{\pm\omega \mathbf{k}_\perp} = \int_{-\infty}^{+\infty}   \frac{dk_z}{\sqrt{2 \pi a k_0} } e^{\pm i \vartheta(k_z)\omega} \phi_{\mathbf{k}}\,,
 \label{expscaunmin_app}
\end{equation}
where $k_0= \sqrt{k_\perp^2+k_z^2}$ is the energy of the photon and where we defined the rapidity $\vartheta(k_z)$ as:
\begin{equation}
   \vartheta(k_z) = \frac{1}{2a} \ln{\frac{k_0+k_z}{k_0-k_z}}\,, 
   \label{rapidity_app}
\end{equation}
and the othornormal Minkowski scalar modes are:
\begin{equation}
    \phi_{\mathbf{k}} = \frac{e^{- i k\cdot x}}{\sqrt{(2 \pi)^3 2 k_0}}\,. 
\end{equation}
(Here, we are using the notation $k^\mu = (k_0,k_z,k_x,k_y)$.)
 Thus, we find
 \begin{equation}
    W_\mu^{(\text{II},\pm, \omega, \mathbf{k}_\perp)} =  i\int_{-\infty}^{+\infty}  \frac{d k_z}{\sqrt{2 \pi a k_0}}e^{\pm i \vartheta(k_z)\omega} \varepsilon_\mu(\mathbf{k}) \phi_{\mathbf{k}}\,,
    \label{bigWexp_app}
\end{equation}
where the polarization vector is given by
\begin{equation}
    \varepsilon^\mu(\mathbf{k}) = \left( \frac{k_z}{k_\perp}, \frac{k_0}{k_\perp},0,0\right)\,,
    \label{polarization_app}
\end{equation}
which satisfies $k\cdot \varepsilon(\mathbf{k})=0$ and  $\varepsilon(\mathbf{k})\cdot \varepsilon(\mathbf{k})=-1$.  This polarization vector is gauge-equivalent to
\begin{equation}
    \widetilde{\varepsilon}_\mu(\mathbf{k}) = \varepsilon_\mu(\mathbf{k}) - \frac{k_z}{k_\perp k_0}k_\mu\,,
\end{equation}
which satisfies $\widetilde{\varepsilon}_t(\mathbf{k}) = 0$ in addition. The relation \eqref{bigWexp_app} between the  Unruh  and  Minkowski  modes  translates  to  that  between  the  Unruh  and Minkowski creation operators as:
\begin{eqnarray}
a^\dagger_{(\text{II}, \pm,\omega, \mathbf{k}_\perp)} =  i\int_{-\infty}^{+\infty}  \frac{d k_z}{\sqrt{2 \pi a k_0}}  e^{\pm i \vartheta(k_z)\omega} b^{\dagger}_{\mathbf{k}}.
\label{operatorexpansionsuppmat_app}
\end{eqnarray}

\section{One-photon interaction probability}
\label{app:C}

In this Appendix, we find the total one-photon emission probability as an integral over the Minkowski momenta $\mathbf{k}$. 
We start from the one-particle part of the final state. It is given by:
\begin{equation}
\begin{aligned}
    \ket{f_{1\textrm{-photon}}} & = \int_0^{+\infty} d\omega\int d^2\mathbf{k}_\perp  \\
    & \quad \times \left[ \mathcal{A}^e_{(\omega,\mathbf{k}_\perp)} a^{R\dagger}_{(\omega,\mathbf{k}_\perp)} + \mathcal{A}^a_{(\omega,-\mathbf{k}_\perp)} 
    a^R_{(\omega,-\mathbf{k}_\perp)}\right]\ket{0_{\text{M}}}\,,
\end{aligned}
\label{1-photon-state_app}
\end{equation}
where the operators $a^R_{(\text{II},\omega,\mathbf{k}_\perp)}$ are written as $a^R_{(\omega,\mathbf{k}_\perp)}$ for simplicity.

By denoting the annihilation operators for the left Rindler modes $A_\mu^{L(\text{II},\omega, \mathbf{k}_\perp)}$ by $a^L_{(\omega,\mathbf{k}_\perp)}$, we can
translate  the relations between the Rindler and Unruh modes, Eq.~\eqref{Unruhmodes}, into those among the creation and annihilation operators as follows:
\begin{equation}
    \begin{aligned}
        a^R_{(\omega,\mathbf{k}_\perp)} & = \frac{a_{(-,\omega,\mathbf{k}_\perp)} + e^{-\pi\omega/a}a^\dagger_{(+,\omega,-\mathbf{k}_\perp)}}
        {\sqrt{1-e^{-2\pi\omega/a}}}\,,\\
        a^L_{(\omega,\mathbf{k}_\perp)} & = \frac{a_{(+,\omega,\mathbf{k}_\perp)} + e^{-\pi\omega/a}a^\dagger_{(-,\omega,-\mathbf{k}_\perp)}}
        {\sqrt{1-e^{-2\pi\omega/a}}}\,,
    \end{aligned}
    \label{RL-as-pm_app}
\end{equation}
where the operators $a_{(\text{II},\pm,\omega,\mathbf{k}_\perp)}$ are written as $a_{(\pm,\omega,\mathbf{k}_\perp)}$ for simplicity. Using
Eqs.~\eqref{1-photon-state_app} and \eqref{RL-as-pm_app}
and the fact that the annihilation operators
$a_{(\pm,\omega,\mathbf{k}_\perp)}$ annihilate the
Minkowski vacuum state $\ket{0_{\text{M}}}$, we find 
\begin{equation}
\begin{aligned}
    \ket{f_{1\textrm{-photon}}} & = \int d^2\mathbf{k}_\perp\int_{0}^{+\infty}d\omega \\ 
& \quad \times \left[\frac{\mathcal{A}^e_{(\omega,\mathbf{k}_\perp)}}{\sqrt{1-e^{-2\pi\omega/a}}}a_{(-,\omega,\mathbf{k}_\perp)}^\dagger \right. \\
& \qquad \quad \left. 
+ \frac{\mathcal{A}^a_{(\omega,-\mathbf{k}_\perp)}}{\sqrt{e^{2\pi\omega/a}-1}}a_{(+,\omega,\mathbf{k}_\perp)}^\dagger \right]\ket{0_{\mathrm{M}}}\,.
\label{app-C:1-photon-Rindler}
\end{aligned}
\end{equation}
Hence, the total one-photon interaction probability is
\begin{equation}
    \begin{aligned}
        P_{\text{tot}} & = \bra{f_{1\textrm{-photon}}} f_{1\textrm{-photon}}\rangle \\
        & = \int_0^{+\infty}d\omega \int d^2\mathbf{k}_\perp
        \left[ \frac{\left|\mathcal{A}^e_{(\omega,\mathbf{k}_\perp)}\right|^2}{1-e^{-2\pi\omega/a}} + 
        \frac{\left|\mathcal{A}^a_{(\omega,-\mathbf{k}_\perp)}\right|^2}{e^{2\pi\omega/a} - 1}\right]\,,
    \end{aligned}
    \label{probsuppmat_app}
\end{equation}
as expected. In order to recover the Larmor formula in the Rindler frame, our goal is to write this probability as a sum over all the momenta. Using Eq.~\eqref{operatorexpansionsuppmat_app}, Eq.~\eqref{1-photon-state_app} can be written as
\begin{equation}
\begin{aligned}
 &  \ket{f_{1\textrm{-photon}}}  \\
 & = i\int d^2\mathbf{k}_\perp\int_0^{+\infty}d\omega
    \int_{-\infty}^{+\infty} \frac{dk_z}{\sqrt{2\pi ak_0}}   \\
    & \quad \times \left[ \frac{e^{-i\vartheta(k_z)\omega}\mathcal{A}^e_{(\omega,\mathbf{k}_\perp)}}{\sqrt{1-e^{-2\pi\omega/a}}} + \frac{e^{i\vartheta(k_z)\omega} 
    \mathcal{A}^a_{(\omega,-\mathbf{k}_\perp)}}{\sqrt{e^{2\pi\omega/a} - 1}}\right]b_\mathbf{k}^\dagger\ket{0_{\text{M}}}\,.
    \label{app-C:1-photon-Minkowski}
\end{aligned}
\end{equation}
Then, the total probability takes the form
\begin{equation}
\begin{aligned}
 P_{\text{tot}}   = \int d^2\mathbf{k}_\perp
    \int_{-\infty}^{+\infty} \frac{dk_z}{2\pi ak_0} 
 \left|\int_{-\infty}^{+\infty}d\omega\frac{e^{-i\vartheta(k_z)\omega}\mathcal{A}^e_{(\omega,\mathbf{k}_\perp)}}{\sqrt{1-e^{-2\pi\omega/a}}}\right|^2\,,
\end{aligned}
\label{firstexpprobasuppmat_app}
\end{equation}
where we used
\begin{equation} 
    \frac{\mathcal{A}^a_{(\omega,-\mathbf{k}_\perp)}}{{\sqrt{e^{2\pi\omega/a} - 1}}}
= \frac{\mathcal{A}^e_{(-\omega,\mathbf{k}_\perp)}}{\sqrt{1-e^{2\pi\omega/a}}}\,.
\end{equation}
We note that Eq.~\eqref{firstexpprobasuppmat_app} can directly be shown to be equivalent to Eq.~\eqref{probsuppmat_app} by noting
$dk_z/ak_0= d\vartheta(k_z)$:
\begin{equation}
    \begin{aligned}
     P_{\text{tot}}  & = \int \frac{d^3\mathbf{k}}{2 \pi a k_0} \left|\int_{-\infty}^{+\infty} d \omega \;  \frac{e^{- i \vartheta(k_z)\omega} \mathcal{A}^e_{(\omega,\mathbf{k}_\perp)}}{\sqrt{1-e^{-2\pi\omega/a}}}\right|^2 \\
      & = \frac{1}{2\pi}\int d^2\mathbf{k}_\perp\; d\omega\; d\omega'\int_{-\infty}^{+\infty}d\vartheta\; e^{- i \vartheta(\omega-\omega')} \\
      & \quad \times \frac{\mathcal{A}^e_{(\omega,\mathbf{k}_\perp)}}{\sqrt{1-e^{-2\pi\omega/a}}} \frac{\mathcal{A}^{e*}_{(\omega',\mathbf{k}_\perp)}}{\sqrt{1-e^{-2\pi\omega'/a}}}\\   
         &= \int_{-\infty}^{+\infty}d\omega \int d^2\mathbf{k}_\perp
        \frac{\left|\mathcal{A}^e_{(\omega,\mathbf{k}_\perp)}\right|^2}{1-e^{-2\pi\omega/a}} \\
         & =  \int_0^{+\infty}d\omega \int d^2\mathbf{k}_\perp
       \left[ \frac{\left|\mathcal{A}^e_{(\omega,\mathbf{k}_\perp)}\right|^2}{1-e^{-2\pi\omega/a}} + 
        \frac{\left|\mathcal{A}^a_{(\omega,-\mathbf{k}_\perp)}\right|^2}{e^{2\pi\omega/a} - 1}\right]\,.
\end{aligned}
\end{equation}


\section{Derivation of Eq.~(28)}
\label{app:D}

In this Appendix, we 
write the amplitude $\mathcal{A}(\mathbf{k})$ 
in terms of $F(\tau)$ instead of its Fourier transform. For this purpose, we make use of the following formula~\cite[Eq.~6.796]{gradshteyn2014table}:
\begin{widetext}
\begin{equation}
 \int_{-\infty}^{+\infty} e^{-i\omega y} e^{\pi\omega/2a}K_{i\omega/a}(z)d\omega = \pi a e^{-iz\sinh ay}.
\end{equation}
Using the definition of $\Tilde{F}(\omega)$, the amplitude is
\begin{equation}
\begin{aligned}
    \mathcal{A}(\mathbf{k})   = q\int_{-\infty}^{+\infty} d\tau\Biggl[ i F(\tau) 
    \sinh a(\vartheta-\tau)e^{-i(k_\perp/a)\sinh a(\vartheta-\tau)}
    - \frac{F''(\tau)}{a k_\perp}\int_{k_\perp/a}^{+\infty}\frac{dz}{z}e^{-iz\sinh a(\vartheta-\tau)}\Biggr]\,.
\end{aligned}
\label{Ak-first-expression_app}
\end{equation}
As it stands, this expression is not convenient for identifying the contribution from the period of uniform acceleration because the first term grows exponentially as a function of $\tau$.  We integrate the first term by parts after writing
\begin{equation}
\begin{aligned}
 iF(\tau)\sinh a(\vartheta-\tau)e^{-i(k_\perp/a)\sinh a(\vartheta-\tau)}
 = \frac{1}{k_\perp}F(\tau)\tanh a(\vartheta-\tau)\frac{d\ }{d\tau}e^{-i(k_\perp/a)\sinh a(\vartheta-\tau)}\,,
\end{aligned}
\end{equation}
and we find
\begin{equation}
    \mathcal{A}(\mathbf{k})  = \frac{q a}{k_\perp}\int_{-\infty}^{+\infty} d\tau\Biggl[ \left( \frac{F(\tau)}{\cosh^2 a(\vartheta-\tau)} - \frac{F'(\tau)}{a} \tanh a (\vartheta-\tau) \right) e^{-i(k_\perp/a)\sinh a(\vartheta-\tau)}
       - \frac{F''(\tau)}{a^2}\int_{k_\perp/a}^{+\infty}\frac{dz}{z}e^{-iz\sinh a(\vartheta-\tau)}\Biggr]\,.
       \label{Awith3terms_app}
\end{equation}
For the term involving $F'(\tau)$ in Eq.~\eqref{Awith3terms_app}, we write
\begin{equation}
 e^{-i(k_\perp/a)\sinh a(\vartheta-\tau)} = i \sinh a(\vartheta-\tau) \int_{k_\perp/a}^{+\infty} dz \; e^{-iz\sinh a(\vartheta-\tau)}  \;, 
\end{equation}
where we assumed a convergence term in the exponent $ \sinh a(\vartheta-\tau) \to \sinh a(\vartheta-\tau)-i \epsilon, \epsilon \to 0^+$. Then, by using the identity,
\begin{equation}
\begin{aligned}
& \int_{-\infty}^{+\infty} d \tau g(\tau) \int_{k_\perp/a}^{+\infty} \frac{d z}{z^n} e^{-iz\sinh a(\vartheta-\tau)}
=  \frac{i}{a}\int_{-\infty}^{+\infty} d \tau \frac{d}{d \tau} \left[ \frac{g(\tau)}{\cosh a(\vartheta-\tau)} \right] \int_{k_\perp/a}^{+\infty} \frac{d z}{z^{n+1}} e^{-iz\sinh a(\vartheta-\tau)} \,,
\end{aligned} 
\label{identity1_app}
\end{equation}
where $g(\tau)$ is a smooth and compactly supported function and $n$ a natural number, we find
\begin{equation}
\begin{aligned}
    \mathcal{A}(\mathbf{k}) & = \frac{q a}{k_\perp}\int_{-\infty}^{+\infty} d\tau\Biggl[  \frac{F(\tau)e^{-i(k_\perp/a)\sinh a(\vartheta-\tau)}}{\cosh^2 a(\vartheta-\tau)} - \frac{1}{a^2} \frac{d}{d \tau}\{ F'(\tau)[1-\tanh^2 a (\vartheta-\tau)]\}\int_{k_\perp/a}^{+\infty}\frac{dz}{z}e^{-iz\sinh a(\vartheta-\tau)}\Biggr]\, \\
    & = \frac{q a}{k_\perp}\int_{-\infty}^{+\infty} d\tau\Biggl[  \frac{F(\tau)e^{-i(k_\perp/a)\sinh a(\vartheta-\tau)}}{\cosh^2 a(\vartheta-\tau)} - \frac{i}{a^3} \frac{d}{d \tau} \left\{ \frac{1}{\cosh a(\vartheta-\tau)} \frac{d}{d \tau} \left[ \frac{F'(\tau)}{\cosh^2 a(\vartheta-\tau)}\right]\right\}\int_{k_\perp/a}^{+\infty}\frac{dz}{z^2}e^{-iz\sinh a(\vartheta-\tau)}\Biggr]\, .
    \end{aligned}
\label{finalA(k)_app}    
\end{equation}
\end{widetext}
The integral of the second term is bounded as
\begin{equation}
 \left| \int_{k_\perp/a}^{+\infty}\frac{dz}{z^2}e^{-iz\sinh a(\vartheta-\tau)} \right|  \leq \int_{k_\perp/a}^{+\infty}\frac{dz}{z^2} = \frac{a}{k_\perp} \;. 
\end{equation}
Then, because the second term in Eq.\eqref{finalA(k)_app} is exponentially decaying as $\left| \vartheta-\tau\right| \to \infty$, it is subdominant if $|\vartheta|<T$. 

\section{Technical details for the derivation of Eq.~(33)}
\label{app:E}

In this Appendix we provide some details omitted in the derivation of Eq.~\eqref{R-per-k-perp}. The square of the amplitude, $|\mathcal{A}(\mathbf{k})|^2$, without the terms coming from the transient effects is proportional to
\begin{widetext}
\begin{equation}
    \begin{aligned}
        I(k_\perp,\vartheta) & \equiv \left| \int_{-\infty}^{+\infty} d\tau\,\frac{F(\tau)e^{-i(k_\perp/a)\sinh a(\vartheta-\tau)}}{\cosh^2 a(\vartheta-\tau)}\right|^2 \\
        & = \int_{-\infty}^{+\infty} d\tau' \int_{-\infty}^{+\infty} d\tau'' \frac{F(\tau')F(\tau'')e^{-i(k_\perp/a)[\sinh a(\vartheta-\tau') - \sinh a(\vartheta-\tau'')]}}
        {\cosh^2 a(\vartheta-\tau')\cosh^2 a(\vartheta -\tau'')}\,,
    \end{aligned}
\end{equation}
We change the integration variables to $\tau=(\tau'+\tau'')/2$ (the average proper time) and $\sigma = \tau'- \tau''$.  Then we find
\begin{equation}
\begin{aligned}
I(k_\perp,\vartheta) & = \int_{-\infty}^{+\infty} d\tau \int_{- \infty}^{+\infty} d \sigma F(\tau+\sigma/2)F(\tau-\sigma/2) 
\frac{e^{ 2i (k_\perp/a)\cosh{a(\vartheta-\tau)}\sinh{a\sigma/2}}}{[\cosh^2{a(\vartheta-\tau)} + \sinh^2{a \sigma/2}]^2}\,.  
\end{aligned}
\label{eq:FF}
\end{equation}
\end{widetext}
For large $T$ 
the integral $I(k_\perp,\vartheta)$ is approximately equal to the expression obtained by limiting the 
integration range for $\tau$ by $|\tau| < T$ and letting $F(\tau+\sigma/2)F(\tau-\sigma/2)=1$ as long as $|\vartheta|<T$ with $||\vartheta| - T| \gg 1/a$. Using this approximation in Eq.~\eqref{eq:FF} we find that the integrand becomes $\tau$-independent after changing the integration variable from $\vartheta$ to 
$\overline{\vartheta} = \vartheta-\tau$, the rapidity in the rest frame of the charge. Then, the $\tau$-integration results in a factor of $2T$ and we obtain Eq.~\eqref{R-per-k-perp} in the main text.

\nocite{*}




\providecommand{\noopsort}[1]{}\providecommand{\singleletter}[1]{#1}%

\end{document}